\documentclass[onecolumn]{webofc}

\usepackage[varg]{txfonts}   
\usepackage{comment}

\begin{document}

\title{
Dimensionality reduction through tensor factorization : application to \textit{ab initio} nuclear physics calculations
}
%
%

\author{\firstname{Mikael} \lastname{Frosini}\inst{1}\fnsep\thanks{\email{mikael.frosini@cea.fr}} \and
        \firstname{Thomas} \lastname{Duguet}\inst{2,3}\fnsep\thanks{\email{thomas.duguet@cea.fr}} 
        \and
        \firstname{Pierre} \lastname{Tamagno}\inst{1}\fnsep\thanks{\email{pierre.tamagno@cea.fr}}
        \and
        \firstname{Lars} \lastname{Zurek}\inst{4,5}\fnsep\thanks{\email{lars.zurek@cea.fr}}
        }

\institute{CEA, DES, IRESNE, DER, SPRC, LEPh, 13115 Saint-Paul-lez-Durance, France
\and
           IRFU, CEA, Université Paris-Saclay, 91191 Gif-sur-Yvette, France 
\and
           KU Leuven, Department of Physics and Astronomy, Instituut vor Kern- en Stralingsfysica, 3001 Leuven, Belgium
\and
CEA, DAM, DIF, 91297 Arpajon, France
\and
Université Paris-Saclay, CEA, Laboratoire Matière en Conditions Extrêmes, 91680 Bruyères-le-Châtel, France
          }

\abstract{%
The construction of predictive models of atomic nuclei from first principles is a challenging (yet necessary) task towards the systematic generation of theoretical predictions (and associated uncertainties) to support nuclear data evaluation. The consistent description of the rich phenomenology of nuclear systems indeed requires the introduction of reductionist approaches that construct nuclei directly from interacting nucleons by solving the associated quantum many-body problem. In this context, so-called \textit{ab initio} methods offer a promising route by deriving controlled (and systematically improvable) approximations both to the inter-nucleon interaction and to the solutions of the many-body problem. From a technical point of view, approximately solving the many-body Schrödinger equation in heavy open-shell systems typically requires the construction and contraction of large mode-4 (mode-6) tensors that need to be stored repeatedly. Recently, a new dimensionality reduction method based on randomized singular value decomposition has been introduced to reduce the numerical cost of many-body perturbation theory.
This work applies this lightweight formalism to the study of the Germanium isotopic chain, where standard approaches would be too expansive to run. Inclusion of triaxiality is found to improve the overall agreement with experimental data on differential quantities. 
}
\maketitle
\section{Introduction}
\label{intro}
Recent years have witnessed a rapid extension of the range of applicability of first-principle calculations of atomic nuclei. So-called nuclear \textit{ab initio} methods construct models of nuclear systems directly from $A$ structure-less interacting nucleons where
\begin{itemize}
    \item The interaction between nucleons (stemming from residual forces between quarks and gluons) is expanded systematically through Effective Field Theories (EFT)~\cite{Entem2002} in coherence with the underlying Quantum Chromo-Dynamics (QCD)
    \item The corresponding many-body Schrödinger equation is solved to a given accuracy level.
\end{itemize}
Even if the \textit{ab initio} scheme is as of today only partially realized in practice~\cite{Machleidt_2023}, recent advances in terms of attainable mass range~\cite{Miyagi2021,Hu22}, observables~\cite{Yao18,Porro:2024tzt}, phenomenology~\cite{Frosini22b,sun2024} but also uncertainty quantification~\cite{Ekstrom19a} make it the most promising route towards systematic predictions for nuclear data applications. 

The extension of the range of \textit{ab initio} techniques to medium mass and heavier systems is mainly due to the development of expansion many-body methods~\cite{Hergert:2020bxy,Dickhoff2004,Hergert16a,Hagen2010,tichai2020many} where the solutions of the $A$-body Schrödinger equation are written as truncated expansions in order to trade the exponential cost of solving the Schrödinger equation exactly in a given space for a polynomial one. 
Writing the problem in the second quantization language allows to reexpress it in terms of tensors and tensor networks~\cite{tichai2019tf}. 

Under certain approximations related to the truncations of the aforementioned expansions, the mode of these many-body tensors can be kept relatively low (4, 6). However, their dimension \(N\), related to the size of the one-body Hilbert space, can range from 1000 to 4000 in typical calculations of heavy systems. 
Needless to say, manipulating tensors of size \(N^4\) is impossible without further assumptions. 
In weakly correlated nuclear systems, reference states usually preserve many symmetries such that these tensors show sparsity patterns that allow to effectively evaluate the tensor networks. 
However, this is not the case anymore in strongly correlated systems (accounting for the vast majority of nuclei) as these tensors become dense. 

A simple and adaptive method to factorize and build low-rank approximations to these large tensors without having to store the original tensors in memory was introduced recently~\cite{frosini24}. 
In this work, the method is applied along the Germanium isotopic chain to calculate second order perturbation theory correlation energies of deformed superfluid nuclear systems.

Notations and main elements of the formalism are briefly recalled in Sec~\ref{sec-1}.
In Sec.~\ref{sec-2}, the method is applied to Germanium isotopes. 
Conclusion and perspectives are given in Sec.~\ref{sec-3}.

\section{Formalism}
\label{sec-1}

Notations closely follow those introduced in ~\cite{frosini24} and only main expressions are recalled.

\subsection{A-body problem}

\textit{Ab initio} nuclear structure theory aims at finding the eigenstates of the nuclear Hamiltonian \(H\). 
Given a basis of the one-body Hilbert space and the corresponding set of anti-commuting particle creators and annihilators \(\{c,c^\dag\}\), \(H\) is expressed in terms of its 1-body, 2-body, 3-body, etc.\ components as
\begin{equation}
    H\equiv \frac1{(1!)^2}\sum_{\alpha\beta} t_{\alpha\beta}c^\dag_\alpha c_\beta
    +
    \frac1{(2!)^2}\sum_{\alpha\beta\gamma\delta} v_{\alpha\beta\gamma\delta}c^\dag_\alpha c^\dag_\beta c_\delta c_\gamma
    +
    \frac1{(3!)^2}\sum_{\alpha\beta\gamma\delta\zeta\epsilon} w_{\alpha\beta\gamma\delta\zeta\epsilon}c^\dag_\alpha c^\dag_\beta  c^\dag_\gamma c_\epsilon  c_\zeta  c_\delta
    +\cdots,
\end{equation}
where $v_{\alpha\beta\gamma\delta}$ ($w_{\alpha\beta\gamma\delta\zeta\epsilon}$) is anti-symmetric under the exchange of the first or last two (any pair among the first or last three) indices. 
For simplicity, we restrict the derivations to 2-body terms and consider only the mode-4 interaction tensor \(v\) in the rest of the document. 
Three-body interactions are included following the rank reduction procedure~\cite{frosini24}, generating a small and controllable error. 

In practice, the one-body basis is truncated to a finite number of elements \(N\). 
Naively, the storage cost of \(v\) is \(N^4\), but symmetries (in particular spherical symmetry) of the nuclear Hamiltonian allow for a compressed representation~\cite{tichai2021adg}. 

\subsection{BMBPT(2)}

At the heart of expansion methods lies the idea of partitioning the original problem (replacing \(H\) by the grand potential \(\Omega\equiv H-\lambda A\), see ~\cite{Tichai18BMBPT} for details)
\begin{equation}
    \Omega |\Psi^A_\mu\rangle = (E_\mu^A-\lambda A)|\Psi^\mu_A\rangle 
\end{equation}
into a simple ``unperturbed'' part \(\Omega_0\), which is easily solvable (associated to a \textit{well-chosen} reference state {\(|\Phi\rangle\)}), and a residual interaction \(\Omega_1\) treated approximately, such that \(\Omega\equiv\Omega_0+\Omega_1\). 
In the special case of open-shell systems, strong correlations require to use reference states that already capture long-range correlations associated to deformation and superfluidity, as these features are poorly captured by the expansion itself. 
This can be done by authorizing the reference state to break rotational and particle number symmetries. 

In the case of canonical BMBPT(2), the reference state \(|\Phi\rangle\) is taken to be a deformed Bogoliubov product state~\cite{scalesi24} solution of the Hartree-Fock-Bogoliubov equations. 
In particular \(|\Phi\rangle\) is characterized by a set of quasiparticle {(qp)} creation/annihilation operators \(\{\beta,\beta^\dag\}\) that are obtained particle creators and annihilators \(\{c,c^\dag\}\) via a unitary transformation
\begin{equation}
    \begin{pmatrix}
        \beta\\
        \beta^\dag
    \end{pmatrix}
    \equiv
    \begin{pmatrix}
        U & V \\
        V^* & U^*
    \end{pmatrix}
    \begin{pmatrix}
        c \\c^\dag
    \end{pmatrix} \, .
\end{equation}
\(|\Phi\rangle\) is annihilated by {any} \(\beta\). In the quasi-particle basis, \(\Omega_0-\Omega^{00}\) (where $\Omega^{00}\equiv \langle \Phi| H| \Phi\rangle$) is a diagonal 1-body operator with strictly positive entries \(E_{k_i}\) called quasiparticle energies. 
\(\Omega_1\) is a pure 2-body operator
\begin{equation}
    \Omega_1 \equiv H^{22} + H^{31} + H^{13} + H^{40} + H^{04} \, , \label{OmegaQP}
\end{equation}
where, \textit{e.g.}, \(H^{ij}\) contains \(i\) qp creators and \(j\) qp annihilators and is represented by the mode-\((i+j)\) tensor $H^{ij}_{k_1\cdots k_{i+j}}$ expressed in the quasi-particle basis. For example, the component $H^{40}$ is represented by the fully-antisymmetric mode-4 tensor $H^{40}_{k_1k_2k_3k_4}$ according to
\begin{equation}
    H^{40} \equiv \frac{1}{4!} \sum_{k_1k_2k_3k_4} H^{40}_{k_1k_2k_3k_4} \beta^\dag_{k_1}\beta^\dag_{k_2}\beta^\dag_{k_3} \beta^\dag_{k_4} \, . \label{H31qp}
\end{equation}
In particular, its expression directly derives from the antisymmetrised product of \(U\) and \(V\) matrices on the original \(H\)
\begin{equation}
    H^{40}_{k_1k_2k_3k_4} = -  \mathcal A\left( \sum_{\alpha\beta\gamma\delta}
    H_{\alpha\beta\gamma\delta} 
    U^*_{\alpha k_1}
    U^*_{\beta k_2}
    V^*_{\gamma k_3}
    V^*_{\delta k_4}
\right).
\end{equation}
where \(\mathcal A\) acts as an antisymmetrisation operator. 

First-order perturbative amplitudes $C^{40}_{k_1k_2k_3k_4}(2)$ in configuration space and second order energy correction $e^{(2)}$ read
\begin{subequations}\label{eq:pt2}
\begin{align}
    C^{40}_{k_1k_2k_3k_4}(2) &\equiv - \frac{ H^{40}_{k_1k_2k_3k_4}}{E_{k_1}+E_{k_2}+E_{k_3}+E_{k_4}} , \\
    e^{(2)} &\equiv - \frac{1}{4!} \sum_{k_1k_2k_3k_4} \frac{ |H^{40}_{k_1k_2k_3k_4}|^2}{E_{k_1}+E_{k_2}+E_{k_3}+E_{k_4}} .
\end{align}
\end{subequations}

\subsection{Tensor decomposition}

\subsubsection{Denominators}

The mode-4 tensor $D^{40}_{k_1k_2k_3k_4}\equiv (E_{k_1}+E_{k_2}+E_{k_3}+E_{k_4})^{-1}$ capturing the energy denominator in  Eq.~\eqref{eq:pt2} is factorized using the Laplace transform~\cite{Braess05,tichai2019tf}
\begin{equation}\label{eq:decom}
    \tilde{D}^{40}_{k_1k_2k_3k_4} \equiv \sum_{i=1}^{n_\text{d}} d^i_{k_1} d^i_{k_2} d^i_{k_3} d^i_{k_4} \, ,
\end{equation}
which, for a small number \(n_\text{d}\approx10\) of vectors \(d^i\), is essentially exact.

\subsubsection{Hamiltonian}

Computing the BMBPT wave-function thus requires to compute and store fully the antisymmetric tensor \(H^{40}\) of dimensions \(N^4\). 
While enforcing symmetries on \(|\Phi\rangle\) generates sparsity patterns in \(H^{40}\), this is not true for a fully general reference state \(|\Phi\rangle\). In that case, \(H^{40}\) is dense and cannot be constructed. For the latter case, a hierarchical decomposition based on Randomized Singular Value Decomposition (RSVD) was introduced in ~\cite{frosini24}\footnote{For simplification, only the truncated RSVD based decomposition is shown here, see~\cite{frosini24} for further exploiting symmetry and antisymmetry of the original mode-4 tensor.} in order to factorize \(H^{40}\) following
\begin{equation}\label{eq:dham}
    \tilde H^{40}_{k_1k_2k_3k_4} \equiv \sum_\mu s_\mu F_{k_1k_2}^\mu G_{k_3k_4}^\mu.
\end{equation}
Key feature is the possibility to perform this decomposition with no need to compute the full tensor in the first place. 
RSVD was implemented based on a blocked-Lanczos method, with a stochastic estimator of the approximation error, such that the expected error can be monitored during the decomposition.

\subsubsection{Second-order energy}

Based on Eqs.~\eqref{eq:decom} and~\eqref{eq:dham}, a factorized expression for the second order energy $E^{(2)}$ is obtained as
\begin{equation}
    E^{(2)} = E^{(0)} + \sum_{\mu\nu i} s_\mu s_\nu
    \left(
    \sum_{k_1k_2} d^i_{k_1} d^i_{k_2}F^\mu_{k_1k_2}F^\nu_{k_1k_2}
    \right)
    \left(
    \sum_{k_3k_4} d^i_{k_3} d^i_{k_4}G^\mu_{k_3k_4}G^\nu_{k_3k_4}
    \right) .
\end{equation}

\section{Application to the Germanium isotopic chain}
\label{sec-2}

In~\cite{frosini24}, the decomposition was proven to be particularily useful in nuclei where many symmetries are broken simultaneously. 
In that specific case, no {\it a priori} argument can be used to reduce the dimensionality of the problem, while tensor decomposition is shown to be largely insensitive to the underlying symmetries. 

The Germanium isotopic chain hosts a large variety of collective phenomena emerging spontaneously by breaking rotational and / or particle-number symmetry in the ground state \(|\Phi\rangle\), and standard perturbation theory might become untractable. 
It is therefore the ideal playground for dimensionality reduction techniques.

\subsection{Numerical settings}

Calculations are performed in a spherical harmonic oscillator basis with frequency $\hbar\omega=12$~MeV truncated to $13$ major shells ($N=1820$). The representation of three-body operators is restricted by employing three-body states up to $e_{\!_{\;\text{3max}}}=24$.

Calculations are performed using the EM1.8/2.0 Hamiltonian~\cite{Hebeler11a} containing  two- (2N) and three-nucleon (3N) interactions. 
Matrix elements were generated using the Nuhamil code~\cite{Miyagi2023}. 
The three-body force is approximated via the rank-reduction method developed in Ref.~\cite{Frosini21a}.

\subsection{Mean-field properties}

\begin{figure}
\centering
\includegraphics[width=\textwidth,clip]{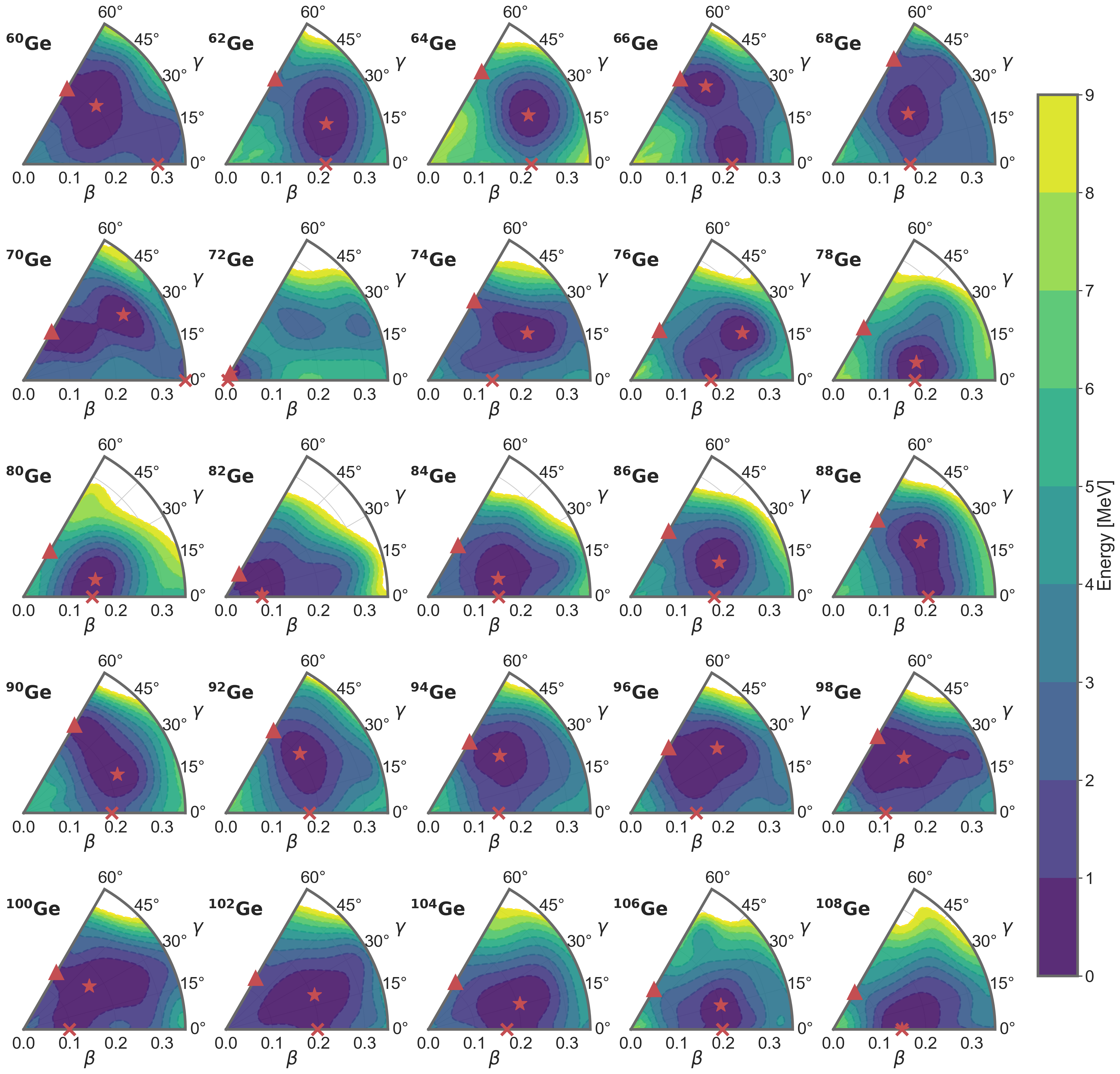}
\caption{HFB total energy surfaces of even-even Germanium isotopes ranging from ${}^{60}\text{Ge}$ to ${}^{108}\text{Ge}$ in the $(\beta,\gamma)$ deformation plane. Prolate, oblate and triaxial
mimina of the energy surfaces are marked with red crosses, triangles and stars, respectively.}
\label{fig:pes}       
\end{figure}

This study focuses on even-even Germanium isotopes ranging from mass $A=60$ to $A=108$. 
Fig.~\ref{fig:pes} displays the triaxial mean-field energy surfaces over the whole isotopic chain. 
Prolate, oblate and triaxial minima are highlighted with red crosses, triangles and stars, respectively. 
It appears that nearly all Germanium are predicted to be deformed at the mean-field level, a vast majority of them being even triaxial with $\gamma\in[15^\circ,45^\circ]$. 
Spherical or near-spherical shape is predicted for $A=72$ and\footnote{
Experimental data~\cite{Zamrun2010} seems to indicate that the shape of $^{72}\text{Ge}$ is not be spherical, but is rather a superposition of deformed and spherical shapes, in agreement with predictions made using phenomenological Gogny interaction~\cite{gogny}. 
This notably explains the deficiencies in the region observed in Fig.~\ref{fig:energies}. } $A=82$.
Some isotopes ($A=66,70,72,76,88$) 
shows signs of shape coexistence, presenting distinct minima close in energy but separated by a saddle point. 
While this is not the topic of this study, this suggests that these nuclei would be an ideal testing ground for multi-reference methods in a future study.

\begin{figure}
\centering
\includegraphics[width=\textwidth,clip]{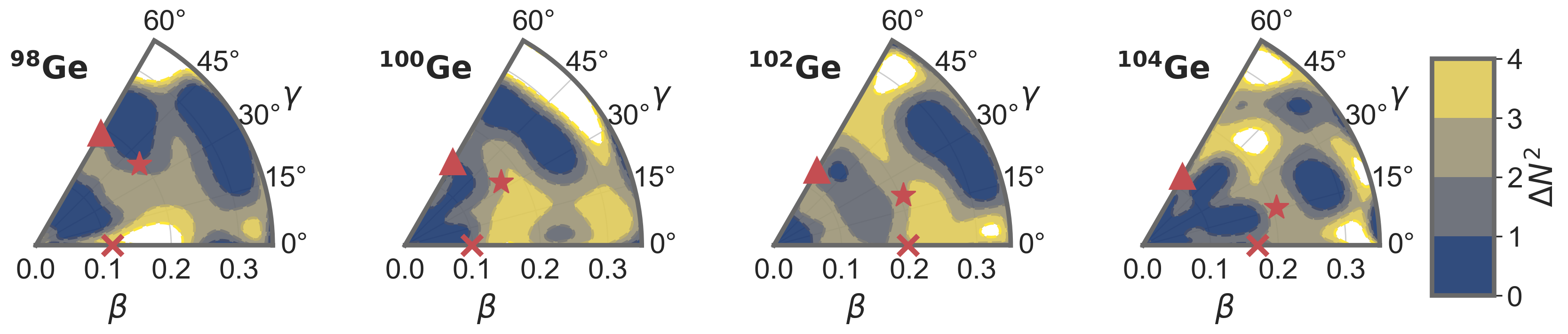}
\caption{HFB neutron number variance of ${}^{98,100,102,104}\text{Ge}$ in the $(\beta,\gamma)$ deformation plane. Prolate, oblate and triaxial
mimina of the energy surfaces are marked with red crosses, triangles and stars, respectively.}
\label{fig:pairing}       
\end{figure}

\begin{figure}
\centering
\includegraphics[width=.7\textwidth,clip]{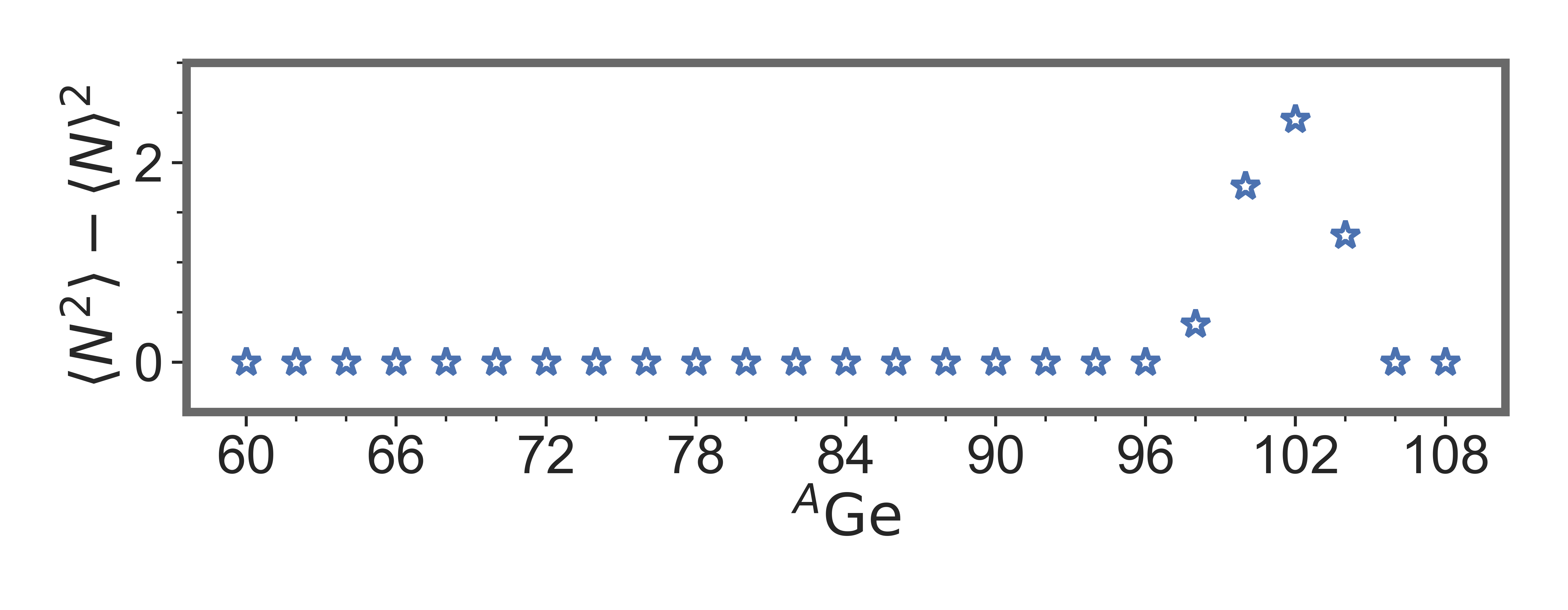}
\caption{HFB neutron number variance of the triaxial minimum along the Ge isotopic chain.}
\label{fig:pairchain}       
\end{figure}

Similarly to results presented in~\cite{frosini24}, most absolute minima are found to be unpaired and do not spontaneously break particle number symmetry, allowing to drastically reduce the cost of the perturbative expansion. 
However, isotopes for $A=98,100,102,104$ are in fact breaking particle number symmetry. 
This can be inferred from Fig.~\ref{fig:pairing}, where neutron number variance is displayed in the deformation plane. 
In all four cases, the triaxial (absolute) minimum is shown to have significant variance translating as the emergence of superfluid correlations. 
In these examples, symmetry-based dimensionality reduction techniques cannot be applied, and tensor-factorized BMBPT is drastically reducing the dimensionality of the problem. 
In order to check the robustness of this observation, the neutron number variance is shown for the triaxial absolute minimum as a function of the nuclear system in Fig~\ref{fig:pairchain}. 
It is striking that neutron number symmetry breaks spontaneously only in the vicinity of ${}^{102}\text{Ge}$, although understanding the reason behind this sudden increase is left for a future work.

\subsection{(B)MBPT calculation}

\begin{figure}
\centering
\includegraphics[width=.8\textwidth,clip]{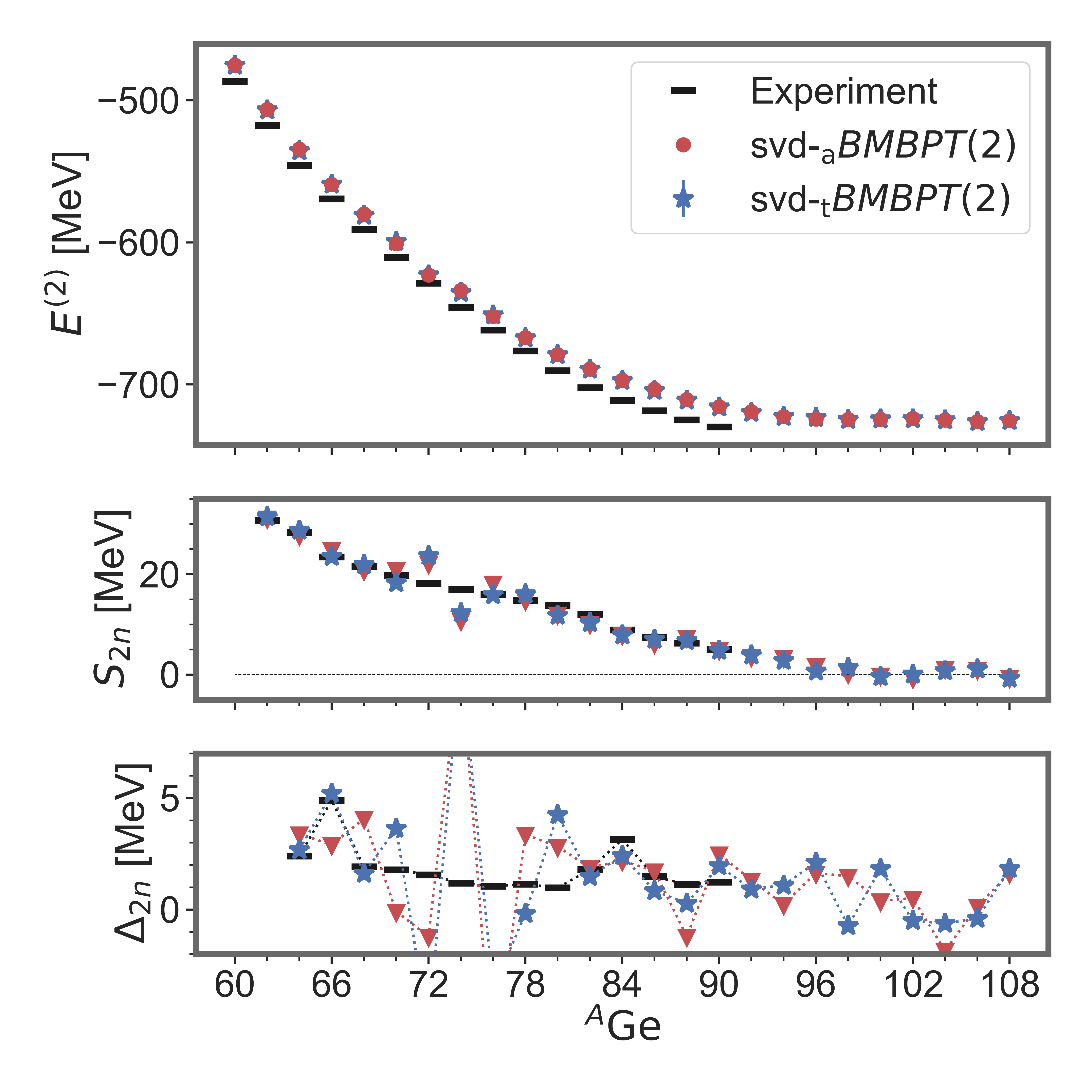}
\caption{Systematic calculations along the Ge isotopic chain. Top panel : $\text{svd-}BMBPT(2)$ binding energy prediction. 
Middle (lower) panel : two neutron separation energies (two-neutron shell gaps). Experiment is indicated with black bars, $_\text{a}BMBPT(2)$ with red triangles and $_\text{t}BMBPT(2)$ with blue stars. 
In the last panel, lines connecting the points are added to guide the eye.}
\label{fig:energies}       
\end{figure}

The second-order prediction of the binding energy is calculated with an estimated error threshold from the truncated SVD of 0.1\% in axial and triaxial systems. 
The upper panel of Fig.~\ref{fig:energies} shows the binding energy at second order as a function of the system, both for the optimal axial HFB state (comparing prolate, oblate and spherical minima) and for the global triaxial minimum identified in Fig.~\ref{fig:pes}. 
While the SVD introduces a controlled truncation error, this error would not be visible on the scale of the figure and is therefore not reported\footnote{In particular, this error is small compared to the one stemming from the truncation of the (B)MBPT expansion~\cite{frosini24,scalesi24}.}.

In order to compare theoretical prediction with experiment, differential quantities (two neutron separation energies $S_{2n}$ and two-neutron shell-gaps $\Delta_{2n}$) are introduced as~\cite{scalesi24}
\begin{subequations}
    \begin{align}
        S_{2n}(N,Z) &\equiv E(N+2,Z) - E(N,Z),\\
        \Delta_{2n}(N,Z) &\equiv E(N+2,Z) + E(N-2,Z) - 2E(N,Z).
    \end{align}
\end{subequations}
$S_{2n}$ and $\Delta_{2n}$ are respectively displayed in the middle and lower panel of Fig~\ref{fig:energies}. 
A good agreement with experimental two-neutron separation energies is found for both axial and triaxial reference states except around $A=72$. 
The discrepancy at $A=72$ is identified as stemming from the predicted spherical shape that leads to an overestimation of the ${}^{72}\text{Ge}$ binding energy.
This correspondingly affects two-neutron shell gaps that are badly reproduced around $A=72$. 
However, the triaxial $\text{svd-}_{\text{t}}BMBPT(2)$ is shown to be significantly more accurate than the axial $\text{svd-}_{\text{a}}BMBPT(2)$ on the rest of the chain, predicting accurately the experimental trends near $A=66,84,88$, while the axial version struggles predicting the changes in curvature.

While this effect is expected to become less significant at higher perturbation orders or in non-perturbative expansion methods, the benefit of triaxiality is still present at second order and improves the agreement with experiment. 

\section{Conclusion}
\label{sec-3}
This work applies the recently derived $\text{svd-}BMBPT(2)$ formalism to perform calculations from first principles along the Germanium isotopic chain. 
This chain is found to host a large variety of collective phenomena (triaxiality, superfluidity) which would prevent applying usual formalisms that require computing and storing large tensors that largely exceed current numerical limitations. 
Accepting a controlled error of $0.1\%$ on ground state energies, experimental binding energies and differential quantities are well reproduced, except in the vicinity of ${}^{72}\text{Ge}$ where a subshell closure is wrongly predicted by our method. 
The inclusion of triaxility in the reference state (whenever axiality is spontaneously broken at the mean-field level) is shown to have a small but significant impact that positively affects differential quantities like two-neutron shell gaps. 
As the dominant source of errors stems from the truncation of the perturbative expansion itself, the next steps will consist in extending the formalism to third order to get a more accurate prediction of drip-line properties. 
Systematic calculations of systems displaying strong correlations are also
envisioned in the near future.

\section*{Acknowledgement}
 The work of M.F. and P.T. was supported by the CEA-SINET project. This work was performed using CCRT HPC resources (TOPAZE supercomputer).

\bibliography{bibliography}

\end{document}